\documentclass[12pt,a4paper]{article}
\usepackage[]{amsmath,amssymb}
\usepackage{graphics,epsfig}
\usepackage[latin1]{inputenc}
\usepackage[T1]{fontenc}
\usepackage{amsmath}
\usepackage{amsfonts}
\usepackage{amssymb}
\usepackage{latexsym}
\usepackage[english]{babel}

\newcommand{\be}{\begin{equation}}
\newcommand{\ee}{\end{equation}}
\newcommand{\beq}{\begin{equation}}
\newcommand{\eeq}{\end{equation}}
\newcommand{\bea}{\begin{eqnarray}}
\newcommand{\eea}{\end{eqnarray}}

\def\be{\begin{equation}}
\def\ee{\end{equation}}
\def\ba{\begin{eqnarray}}
\def\ea{\end{eqnarray}}

\begin{document}
\title{Remarks on entanglement entropy for gauge fields}
\author{Horacio Casini\footnote{e-mail: casini@cab.cnea.gov.ar}, Marina Huerta\footnote{e-mail: marina.huerta@cab.cnea.gov.ar} and Jos\'e Alejandro Rosabal\footnote{e-mail: alejo$\_$rosabal@yahoo.es}\\{\sl Centro At\'omico Bariloche,
8400-S.C. de Bariloche, R\'{\i}o Negro, Argentina}}
\date{}
\maketitle

\begin{abstract}
In gauge theories the presence of constraints can obstruct expressing the global Hilbert space as a tensor 
product of the Hilbert spaces corresponding to degrees of freedom localized in complementary regions. In 
algebraic terms, this is due to the presence of a center --- a set of operators which commute with all 
others --- in the gauge invariant operator algebra corresponding to finite region. A unique entropy can be assigned to 
algebras with center, giving place to a local entropy in lattice gauge theories. However, ambiguities 
arise on the correspondence between algebras and regions. In particular, it is always possible to choose 
(in many different ways) local algebras with trivial center, and hence a genuine entanglement entropy, 
for any region. These choices are in correspondence with maximal trees of links on the boundary, which 
can be interpreted as partial gauge fixings. This interpretation entails a gauge fixing dependence of 
the entanglement entropy. In the continuum limit however, ambiguities in the entropy are given by terms local 
on the boundary of the region, in such a way relative 
entropy and mutual information are finite, universal, and gauge independent quantities. 
\end{abstract}

\section{Introduction}

The standard procedure to compute the entropy contained in some region $V$ in an extended system  
requires to express the global Hilbert space as a tensor product of the Hilbert spaces ${\cal H}_V$ generated by the 
degree of freedom in $V$ and the Hilbert space ${\cal H}_{\bar{V}}$ generated by the degrees of freedom in the complementary region $\bar{V}$. 
The reduced state in $V$ is given by 
 \begin{equation}
 \rho_V=\textrm{tr}_{{\cal H}_{\bar{V}}}(\rho)\,,\label{fomi}
 \end{equation}
 where $\rho$ is the global state. 
  This is the only state in ${\cal H}_V$ giving the correct expectation values for all local operators in $V$:
 \begin{equation}
 \textrm{tr}(\rho {\cal O}_V)=\textrm{tr}(\rho_V {\cal O}_V)\,.
 \end{equation}    
Hence, local entropy is defined as the von Neumann entropy of $\rho_V$
\begin{equation}
S(V)=-\textrm{tr}(\rho_V \log \rho_V)\,. \label{fomi1}
\end{equation}
If the global state is pure this is the entanglement entropy of the bipartition ${\cal H}_V\otimes {\cal H}_{\bar{V}}$, and measures 
in a precise operational sense the degree of entanglement between $V$ and $\bar{V}$ \cite{nielsen}. In the case of impure global states, $S(V)$ 
is more generally the entropy in the region $V$, and also contains for example the thermal entropy.   

The quantities with most direct physical significance for extended systems are the local operators and 
the global state, which produces the expectation values. 
These are also the basic elements for the continuum QFT limit. 
In the above construction of the local entropy $S(V)$, however, we are forced to consider the local Hilbert spaces ${\cal H}_V$, where the 
local operators are linearly represented. These Hilbert spaces are less direct quantities and, as we will discuss in this paper, they may be considered specially unnatural 
for lattice gauge theories due to the presence of constraints. Depending on the details, constraints can impede the interpretation of the state reduction to a region in terms of tensor products. 

Entanglement entropy for gauge fields has been considered previously in the literature \cite{Kabat,fursaev,Wall,Solodukhin,Zhitnitsky,Iellici:1996jx,Dowker,theisen,headrick,Polikarpov,donnelly}, and often being related to puzzling results. 
 In relation to black 
hole entropy there is the early work by Kabat, where he found a negative contact term \cite{Kabat}. This was followed by several interpretations 
(e.g. \cite{Wall,Solodukhin,Zhitnitsky,Iellici:1996jx}). 
Within field theory calculations, a mismatch on the logarithmic coefficient in the entanglement entropy with respect to the expected anomaly coefficient was reported \cite{Dowker,theisen}.
In the lattice, difficulties in expressing the global Hilbert space as a tensor product have been understood as a consequence of the fact that elementary excitations 
in gauge fields are associated to closed loops rather than points in space \cite{Polikarpov,donnelly}. In these works, it was argued the Hilbert space has to be extended to properly define an entanglement entropy. The result in the extended space contains a quantum bulk contribution and a boundary classical Shannon term. 
These subtleties found in the lattice formulation may be related to the continuum issues but their 
relation is not clearly established so far and still calls for a deeper understanding \cite{headrick}.  

In this paper, as a first step on this direction, we focus on lattice gauge fields, and we take the discussion into a broader context within an algebraic approach.

In the next section we introduce lattice gauge fields and describe the local generators for the gauge invariant operator algebra. In section 3 we review how a unique entropy can be computed for a state acting on an algebra, giving place to a local gauge invariant entropy for gauge fields. 

The problem of the identification of degrees of freedom with regions
is translated to the one of assignations of algebras to regions. We show there is no unique choice and discuss several possibilities in section 4. 
Some natural geometric choices for local algebras contain a non trivial center, preventing the interpretation of local 
entropy as entanglement entropy. 

In this scenario, the prescription introduced in the literature to solve the puzzle of 
localization of degrees of freedom within a region \cite{Polikarpov,donnelly}, corresponds to a particular  choice of algebra, we called \textit{the electric center choice}. 
However, we  show  that some minor modifications in the choice of algebra for 
a given region lead to local algebras with trivial center, and hence to a tensor product interpretation and an entanglement entropy.  The local algebras with trivial 
center are in correspondence with  maximal trees of boundary links. The partial gauge fixing induced by the boundary maximal tree, 
tells us the ambiguities can be interpreted as a \textit{gauge fixing dependence} in the entropy. 

Hence, in the case of gauge theories the presence of a center for the most natural choices of local algebra render manifest the ambiguities 
inherent to the {\sl relation between operator algebras and regions}.  
These ambiguities also tarnish the case of other fields (e.g. a scalar field), and in a certain sense they are more closely related to the idea 
of defining a geometric region  in a regulated geometry (such as a lattice) using only the physical content of the model, than to the peculiar 
properties of gauge theories.  In section 5 we will see these ambiguities produce typically large numerical 
variations in the value of the entropy, while they are quite harmless for the relative entropy quantities. 
Along the same line, in section 6 we will discuss how the continuum limit removes conceptual differences between entanglement entropy for gauge theories with the one 
for other kind of fields. Universal information in the entropy is independent of the ambiguities on the choice of algebra. We end with some comments in section 7. 

\section{Lattice gauge fields}
The basic variables for gauge fields in a lattice\footnote{For a review see for example \cite{ks}} (at fixed time) are elements $U_{(ab)}\in G$ of the gauge group $G$ assigned to each oriented 
link $l=(ab)$ joining lattice vertices $a,b$. The link $\bar{l}=(ba)$ with the reverse orientation has assigned the 
inverse group element $U_{\bar{l}}=U_{(ba)}=U_{(ab)}^{-1}=U_l^{-1}$. The variables $g_a$ of the gauge transformations 
are also elements of the group $ G$ but they are attached to the vertices $a=1,...,N_V$ of the lattice.  The gauge transformation 
law is $U_{(ab)}^\prime=g_a U_{(ab)} g_b^{-1}$. 

Consider the vector space ${\cal V}$ of all complex wave functionals $|\Psi\rangle\equiv \Psi[U]$, where $U=\{U_{(ab)}\}$ 
is an assignation of group elements to all links. The wave functionals describing actual physical states form the 
subspace ${\cal H}\subset {\cal V}$ of gauge invariant functionals, 
\begin{equation}
\Psi[U]=\Psi[U^g]\,,
\end{equation}    
where $U^g=\{g_a U_{(ab)}g_b^{-1}\}$. The scalar product is defined in ${\cal V}$ as
\begin{equation}
\langle \Psi_1|\Psi_2\rangle=\sum_{U_{1}} ... \sum_{U_{N_L}} \Psi_1[U]^* \Psi_2[U]\label{scalar}
\end{equation}
where $U_l$ is the variable corresponding to the link $l=1,...,N_L$, for a lattice with $N_L$ links. That is, 
the scalar product is defined by an orthogonal basis given by the characteristic functions on 
the different configurations of the variables on links (i.e. functions which are $1$ on some configuration 
and zero for all other configurations). For compact continuum groups the sum over elements of the group is replaced 
by integration over the gauge group with the invariant Haar measure. The subspace ${\cal H}$ of gauge 
invariant functions also forms a Hilbert space with the scalar product (\ref{scalar}). 

The algebra of physical operators ${\cal B}({\cal H})$ is the subalgebra of the algebra ${\cal B}({\cal V})$ of all 
linear operators with domain and range in the physical subspace ${\cal H}$. We need to understand the local structure 
of these operators in order to construct local algebras of operators assigned to space-time regions. To keep the 
discussion of the next sections as simple as possible and avoid complications which could obscure the 
main arguments, from here on we restrict ourselves to the case of abelian gauge groups. The discussion will also be focused on finite groups of dimension $d_G$, and eventually we add some explanations for the case of continuous group.

A set of generators for the algebra of all (gauge and non gauge invariant) operators ${\cal B}({\cal V})$ is constructed 
in a straightforward way. The space ${\cal V}$ is a tensor product over 
the links of the $d_g$-dimensional complex vector space $C^{d_G}_l$ for a single link $l$. The algebra ${\cal B}({\cal V})$ is the tensor product 
over links of the algebra $GL(C,d_G)_{l}$ of complex $d_G\times d_G$ matrices acting on $C^d_l$. We will first describe a 
complete set of generators for the algebras $GL(C,d_G)_{l}$ on single links, and then analyze how to construct generators 
for the gauge invariant algebra. 

A particular example of operators are the unitary operators induced by an element $g$ of $G$ acting 
 on a given link $l$,
\be
(\hat{L}_g^{l} \Psi)[U_1,...,U_N]=\Psi[U_1,...,g U_{l},...,U_n]\,.
\ee
We have $\hat{L}_g^{l}=\hat{L}_{g^{-1}}^{\bar{l}}$, $\hat{L}_{g_1}^l \hat{L}_{g_2}^l=\hat{L}_{g_1 g_2}^l$. Hence, for 
abelian groups, these operators commute to each other for different groups elements and links. 

Therefore, for a single link, the operators $\hat{L}_g^l$ form an abelian algebra. In order to complete the generators for the 
single link algebra, we introduce an analogous of the coordinate operators in the description of the wave function
\be
(\hat{U}^{r}_l \Psi)[U]=U^{r}_l \Psi[U]\,,
\ee
where $U^r_l$ is the numerical value corresponding to $U_l$ in the (one dimensional) representation $r$. The $\hat{U}^r_l$ 
for different $r$ (and $l$) clearly commute. Then we have two commuting algebras $\hat{U}^r_l$ and $\hat{L}_g^l$ for the single 
link vector space, which are analogous to the coordinate and momentum operators for a harmonic oscillator. It is not difficult 
to see they do not commute to each other, and together they are a generating set for the algebra of single link 
operators.\footnote{For example, in the basis of the $d_G$ vectors $\delta_{U^l,g}$ the $\hat{U}^r_l$ span all the diagonal 
matrices. The $\hat{L}_g^l$ for different $g$ can take any basis vector to any other. Any matrix can be generated with linear combinations of these operations.}  

Now, we want to understand how to reduce this set of generators to produce generators for the gauge invariant algebra.  
For abelian gauge groups the operators $\hat{L}^l_g$ are already gauge invariant. That is, if $\Psi[U]$ is gauge invariant, $(\hat{L}_g^l \Psi)[U]$ 
is also gauge invariant. Then, it is only left to see how to make a gauge invariant version of the coordinate operators $\hat{U}^r_l$. 
Let us define an operator induced by the gauge transformation by an element $g$ on the vertex $a$ 
\bea
(\hat{T}_{g_{a}} \Psi)[U]= \Psi[ U^{g_a}]\,,\\
\hat{T}_{g_{a}}=\prod_b \hat{L}_g^{(ab)} \,,
\eea
where the product is over the vertices $b$ connected to $a$ by some link in the lattice.
For a gauge invariant state $\Psi[U]$ we have
\bea
 \left(\hat{T}_{g_a} \hat{U}^{r}_{(ab)}\Psi\right)[U]= g^{r}U^{r}_{(ab)} \Psi[U]\neq \left( \hat{U}^{r}_{(ab)}\Psi\right)[U]\,,\\
\left(\hat{T}_{g_a} \hat{U}^{r}_{(ca)}\Psi\right)[U]= U^{r}_{(ca)} (g^{-1})^{r} \Psi[U]\neq \left( \hat{U}^{r}_{(ca)}\Psi\right)[U]\,.
\eea  
This shows the operators $\hat{U}_l^r$ are not gauge invariant. However, the product of operators $\hat{U}_{(ca)}^{r}\hat{U}_{ab}^{r}$ 
is invariant under $\hat{T}_{g_a}$. Hence, only  products
 of operators in a closed line formed by oriented links are invariant under gauge transformations based on any vertex. These are the 
Wilson loop operators
 \begin{equation}
 \hat{W}^r_\Gamma=\hat{U}^{r}_{(a_1 a_2)}\hat{U}^{r}_{(a_2 a_3)}...\hat{U}^{r}_{(a_k a_1)}\,,
 \end{equation}
 where $\Gamma={a_1 a_2 ... a_k a_1}$ is an oriented closed path made by links in the lattice. 
 
Wilson loop operators all commute with each other (for loop paths on a fixed time as considered in this fixed time Hilbert space description). 
In the Schr\"odinger representation of ordinary quantum mechanics all wave functions can be thought to arise from the identity 
function $\psi(x)=1$ by acting on it with functions of the coordinate operator. Analogously here, all gauge invariant wave functions 
arise from the trivial function $\Psi^0[U]=1$ by acting on it with arbitrary combinations of Wilson loop operators, for different paths $\Gamma$ 
and group representations.

A gauge invariant state satisfies
\be
(\hat{T}_{g_{a}} \Psi)[U]=\Psi[U]\,, \hspace{2cm} \Psi[U]\in {\cal H}\,.
\ee
An operator is gauge invariant (belongs to ${\cal B}({\cal H})$) if it commutes with all $\hat{T}_{g_a}$ (on the physical subspace). Hence, 
on the physical subspace we have the {\sl constraint equations}
\be
\hat{T}_{g_a}=\prod_b \hat{L}_g^{(ab)}\equiv 1\,.\label{14}
\ee
These imply the operators $\hat{L}^l_g$ are not all independent for different links. 

In conclusion, a generating set of operators for the gauge invariant algebra of operators is given by the Wilson loop operators and the link operator $\hat{L}^l_g$.

For continuous groups, the link variables can be parametrized $U_l=e^{i a A_l}$ in terms of the vector potential $A_l$ and lattice spacing $a$. In the continuum, this is assimilated to $A_\mu dx^\mu$, with $dx^\mu$ describing the displacement vector along the link. Wilson loops are defined as above, and link operators are replaced by the electric operator $\hat{E}_l$ which is the conjugate momentum to the coordinate operator $\hat{A}_l$, and does not commute with the (non gauge invariant) operator $\hat{U}_l$, 
\be
[\hat{E}_l,\hat{A}_{l^\prime}]=- i \delta_{l,l^\prime} \,,\hspace{2cm}[\hat{E}_l,\hat{U}_{l^\prime}]=  \hat{U}_l \delta_{l,l^\prime}\,.
\ee
Hence, $\hat{E}_l$ is the generator of translations in the group and plays a role analogous to the finite translations $\hat{L}_g^l$ for finite groups. The constraint equation is the infinitesimal version of (\ref{14})
\be
\sum_{b} \hat{E}_{(ab)}=0\,.
\ee  

\section{Localized entropy in gauge theories}

In this section we analyze the problem of the local entropy in algebraic terms, 
and show that the local entropy for gauge theories (and in fact for any theory) naturally fits into the general definition of an entropy 
associated to a state on an algebra which is discussed elsewhere in the literature \cite{petz}.  

\subsection{Local algebras, constraints and center}

In order to highlight the special features of gauge theories let us first briefly discuss the case of a scalar field on the lattice. For a 
scalar field, the basic variables are the field $\phi(a)$ and momentum $\pi(a)$ for a lattice site $a$. They obey the canonical commutation relations
\begin{equation}
[\phi(a),\pi(b)]=i \delta_{a, b}\,,\hspace{1cm}[\phi(a),\phi(b)]=0\,,\hspace{1cm}[\pi(a),\pi(b)]=0 \,.
\end{equation}
This defines the operator algebras, independently of the election of the state and Hamiltonian. 
For a lattice region $V$ given by a set of sites, the natural algebra ${\cal A}_V$ is the one generated by the set of operators ${\cal G}_V$ 
formed by  $\phi(a),\pi(a)$ for all $a\in V$. That is, ${\cal A}_V$ contains, besides the multiples of the identity, all polynomials of the 
canonical variables localized in $V$.\footnote{This algebra of canonical commutation relations can be completed in appropriate topology to give a $C^*$-algebra. 
The interested reader can consult \cite{brattel}.} Another way to define the algebra generated by a 
set of operators is the following. Given a set of operators ${\cal G}$, we can define its commutant ${\cal G}^\prime$ as the set 
of all operators which commute with all operators in ${\cal G}$. Then, an operator generated by ${\cal G}$ will also commute with all operators 
in the commutant ${\cal G}^\prime$. A well know theorem tells the generated algebra is in fact the double commutant ${\cal G}^{\prime\prime}$ \cite{well}. 
Hence we have
\begin{equation}
{\cal A}_V=\{\phi(a),\pi(a), a \,\textrm{in}\, V\}^{\prime\prime}\,.
\end{equation} 

In the same way we can define the algebra ${\cal A}_{\bar{V}}$ corresponding to the complementary region in the lattice.
Since $\phi(b), \pi(b)$ for $b\notin A$ commute with all $\phi(a),\pi(a)$, $a\in V$, we have ${\cal A}_{\bar{V}}\subseteq ({\cal A}_V)^\prime$, 
and also, if some operator commutes with all the $\phi(a),\pi(a)$ for $a\in A$ then it is generated by the $\phi(b),\pi(b)$ for $b\in \bar{A}$. Hence we have  
\begin{equation}
{\cal A}_{\bar{V}}=({\cal A}_V)^\prime\,,\hspace{2cm} {\cal A}_{V}=({\cal A}_{\bar{V}})^\prime\,.
\end{equation}
In the algebraic approach to QFT this is called Haag's duality \cite{haag}. 

What is more relevant for the present discussion is that since all operators in the global Hilbert space which commute with all the canonical variables are 
proportional to the identity we have
\begin{equation}
{\cal A}_{V}\cap ({\cal A}_{V})^\prime={\bf 1}\,,\label{uno}
\end{equation}
where we have written ${\bf 1}$ for the algebra of operators proportional to the identity. 

The condition (\ref{uno}) allow us to interpret the algebra ${\cal A}_V$ as a {\sl factor} in a tensor product. In fact, a 
tensor product factorization of a Hilbert space ${\cal H}={\cal H}_V\otimes {\cal H}_{\bar{V}}$ is associated to two algebras ${\cal A}_V$ 
and ${\cal A}_{\bar{V}}$ which are formed by the operators of the form $O_V\otimes 1_{\bar{V}}$ and $1_{V}\otimes O_{\bar{V}}$ respectively, 
in such a way ${\cal A}_{V}\cap ({\cal A}_{V})^\prime={\bf 1}=1_V\otimes 1_{\bar{V}}$.

\begin{figure}
\centering
\leavevmode
\epsfysize=4cm
\epsfbox{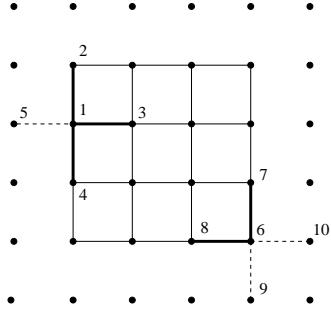}
\bigskip
\caption{
The product of three link operators on the square $\hat{L}_g^{(12)} \hat{L}_g^{(13)} \hat{L}_g^{(14)}$ is equal to a link operator outside the 
square, $\hat{L}_g^{(51)}$, and hence it commutes with the rest of the operators on the square. The same occurs for the 
product $\hat{L}_g^{(67)} \hat{L}_g^{(68)}=\hat{L}_g^{(96)} \hat{L}_g^{(10\,6)}$ on the corner.}
\label{figu1}
\end{figure}

While (\ref{uno}) holds for the local algebras of a scalar field as defined above, the most general situation would rather be that  
\begin{equation}
{\cal A}_V\cap ({\cal A}_V)^\prime={\cal Z}_V\,,
\end{equation}
with a non trivial algebra ${\cal Z}_V$. This is called the {\sl center} of the algebra ${\cal A}_V$ and is just the (mutually commuting) set of operators in the 
algebra which commute with all other. 

The case with non trivial center appears naturally for localized gauge invariant operator algebras in gauge theories. 
Consider for example defining a region $V$ as a subset of the links in the lattice, and the algebra ${\cal A}_V$ as the one generated by the 
Wilson loop operators for paths $\Gamma\subseteq V$ and all link operators $\hat{L}_g^{l}$ for links $l$ in $V$. If there is a link $l\in V$ which does not belong to any loop, the 
algebra contains a center including at least the link operators $\hat{L}^l_g$. Further, the algebra contains a center even in the case all links belong 
to some loop in $V$. This is because of the constraint equation $\hat{T}_{g_a}=\Pi_{(ab)} \hat{L}_g^{(ab)}\equiv 1$, we have that necessarily some  
products of link operators in ${\cal A}_V$ located near the boundary of $V$ are equal to some link operator (or product of link operators) external to $V$. 
For example, in figure \ref{figu1} we have $\hat{L}^{(51)}_g= \hat{L}_g^{(12)} \hat{L}_g^{(13)} \hat{L}_g^{(14)}$. But $\hat{L}_g^{(15)}$ commutes 
with all Wilson loops (and link operators) in $V$. Hence, $\hat{L}_g^{(15)}$ is a non trivial operator in ${\cal A}_V$ which commutes with all other 
operators in ${\cal A}_V$, and the algebra has a non trivial center. Analogously, on the corner of the region in figure \ref{figu1}, we 
have $\hat{L}^{(67)}_g\hat{L}_g^{(68)}=\hat{L}^{(96)}_g\hat{L}_g^{(10\, 6)}$. This product of operators is spatial to the region and 
hence $\hat{L}^{(67)}_g\hat{L}_g^{(68)}$ belongs to the center. 
 
 Though, as we will see, the existence of a non trivial center can be avoided by specific choices of the boundary details in the definition of the algebra, in general, for the most natural choices  of the local algebras constraint 
equations give place to a non trivial center. We will come back to this point in the next section, 
where we look at different options for defining the local algebras. 
 
 Here we emphasize that in the case of non trivial center there is no interpretation of the algebras in terms of tensor product of Hilbert spaces, and 
the usual way of computing the reduced density matrix by a partial trace cannot be implemented. However, there is a well defined notion of entropy 
for a state on a (finite) algebra with center which reduces to the standard formulas (\ref{fomi}) and (\ref{fomi1}) for the case of tensor products 
(or equivalently algebras with trivial center). With this definition we can compute a (gauge invariant) entropy for a state in a region in a lattice 
gauge model as the entropy of the corresponding algebra of gauge invariant operators. More precisely, the meaning of a ``state on an algebra'' is  a 
linear functional on operators with complex values, that is positive for the positive definite operators, and is normalized to one for the unit element. 
In the present context it is more concretely given by the expectation values given by a state in the global Hilbert space restricted to the 
operators in the local algebra.   

\subsection{Operator algebras and entropy}
Hence, let us look at operator algebras with center, and see the fact they do not correspond to algebras in a tensor product of Hilbert spaces is 
not an obstacle for calculating an entropy (see for example \cite{petz}). The idea is first to simultaneously diagonalize all operators in the center, 
which are mutually commuting, and commute with the rest of the algebra. Then, we have the generic element of ${\cal Z}$ writes 
\begin{equation} 
\left(\begin{array}{cccc}
(\lambda^1) & 0 & \hdots & 0 \\
0 & (\lambda^2) & \hdots & 0 \\
\vdots & \vdots & &\vdots \\
0 & 0 & \hdots & (\lambda^m)
\end{array}\right)\,.
\end{equation}
Each of the $(\lambda^k)$ represents  $\lambda^k$ times a unit matrix of some dimension $d_k \times d_k$. The matrices of this form for all 
values $\lambda^k$ span ${\cal Z}$. 

In this diagonalizing basis the rest of operators in the algebra assume a block diagonal form. Since the center ${\cal Z}$ of ${\cal A}$ is also 
the center of the commutant ${\cal A}^\prime$, the elements of ${\cal A}^\prime$ will also take a block diagonal form in the same basis.  
Hence the algebra generated by ${\cal A}$ and ${\cal A}^\prime$ has the general form   
\begin{equation}
{\cal A}{\cal A}^\prime\equiv \left({\cal A}\cup {\cal A}^\prime\right)^{\prime\prime}= 
\left(\begin{array}{cccc}
{\cal A}_1\otimes {\cal A}_1^\prime & 0 & \hdots & 0 \\
0 & {\cal A}_2\otimes {\cal A}_2^\prime & \hdots & 0 \\
\vdots & \vdots & &\vdots \\
0 & 0 & \hdots & {\cal A}_m\otimes {\cal A}_m^\prime 
\end{array}\right)\,.
\end{equation}
The $k^{\textrm{th}}$ block is decomposed as a tensor product of full matrix algebras ${\cal A}_k$ and ${\cal A}_k^\prime$, canonically 
included in ${\cal A}$ and ${\cal A}^\prime$ respectively. The product of the dimensions $b_k$ and $c_k$ of ${\cal A}_k$ and ${\cal A}_k^\prime$ 
is equal to the ones in the center, $b_k c_k=d_k$.  The commutant of the generated algebra $\left({\cal A}\cup {\cal A}^\prime\right)^{\prime\prime}$ 
is the center ${\cal Z}$. Then, if the center is non trivial, ${\cal A}$ and ${\cal A}^\prime$ do not generate all the operators in the global Hilbert space. 

Then, the algebra ${\cal A}$ is isomorphic to the block diagonal representation of full matrix algebras
\begin{equation}
{\cal A}\equiv 
\left(\begin{array}{cccc}
{\cal A}_1 & 0 & \hdots & 0 \\
0 & {\cal A}_2 & \hdots & 0 \\
\vdots & \vdots & &\vdots \\
0 & 0 & \hdots & {\cal A}_m 
\end{array}\right)\,.
\end{equation}
The reduced state on this algebra is defined in the same terms as the reduced density matrix for a tensor product of Hilbert spaces, 
that is, as the unique density matrix {\sl belonging to the algebra} ${\cal A}$ and giving the correct expectation values. More 
explicitly, $\rho_{\cal A}\in {\cal A}$ and $\textrm{tr}(\rho_{\cal A} {\cal O})=\textrm{tr} (\rho {\cal O} )$, for any ${\cal O}\in {\cal A}$, 
where $\rho$ is the global state. 

In the block diagonal representation for the algebra obtaining $\rho_{\cal A}$ from $\rho$ involves two operations. First, the entries of $\rho$ 
which lie out of the blocks are erased such that $\rho_{{\cal A}{\cal A}^\prime}$ belongs to the algebra ${\cal A}{\cal A}^\prime$. These entries do not contribute to the expectation values of operators of ${\cal A}$ or ${\cal A}^\prime$. 
We can write this block diagonal density matrix 
\begin{equation}
\rho_{{\cal A}{\cal A}^\prime}= 
\left(\begin{array}{cccc}
p_1 \rho_{{\cal A}_1{\cal A}_1^\prime} & 0 & \hdots & 0 \\
0 & p_2 \rho_{{\cal A}_2{\cal A}_2^\prime} & \hdots & 0 \\
\vdots & \vdots & &\vdots \\
0 & 0 & \hdots & p_m \rho_{{\cal A}_m{\cal A}_m^\prime} 
\end{array}\right)\,,\label{dua}
\end{equation}
 where $\rho_{{\cal A}_k{\cal A}_k^\prime}$ are density matrices of dimension $d_k\times d_k$ and the positive numbers $p_k$, $\sum_k p_k=1$,    
 guarantee $\textrm{tr}\rho_{{\cal A}{\cal A}^\prime}=1$.
 
 Then, on each block we have to partial trace over the factor ${\cal A}_k^\prime$. The final form of the density matrix $\rho_{\cal A}$ is
\begin{equation}
\rho_{\cal A}= 
\left(\begin{array}{cccc}
p_1 \rho_{{\cal A}_1} & 0 & \hdots & 0 \\
0 & p_2 \rho_{{\cal A}_2} & \hdots & 0 \\
\vdots & \vdots & &\vdots \\
0 & 0 & \hdots & p_m \rho_{{\cal A}_m} 
\end{array}\right)\,,\label{explicit}
\end{equation}
where $\rho_{{\cal A}_k}$ is a density matrix of dimension $b_k\times b_k$, with $\textrm{tr}\rho_k=1$.

The entropy of this density matrix is simply the usual von Neumann entropy
\begin{equation}
S(V)=-\textrm{tr}(\rho_{\cal A}\log \rho_{\cal A})=H(\{p_k\})+\sum_k p_k S(\rho_{{\cal A}_k})\,,\label{er}
\end{equation}
where 
\begin{equation}
H(\{p_k\})=-\sum_k p_k \log(p_k)
\end{equation}
is the classical Shannon entropy of a probability distribution. 
Eq. (\ref{er}) shows the entropy is a sum of the average of the ``entanglement'' part on each sector, plus the classical entropy 
of the probability distribution of the variables on the center. These lasts effectively act as classical commuting variables for the 
algebra, determining superselection sectors to which different classical probabilities are assigned by the global state.  In this 
sense, we remark the entropy of an algebra with center does not have a ``entanglement entropy'' interpretation in the sense given 
to this term in quantum information theory. The entanglement entropy of a reduced state in a bipartite quantum system with global 
pure state is known to be an adequate measure of entanglement, i.e. it measures the number of EPR pairs necessary to form the state 
or that can be distilled from the state with local operations and classical communication \cite{nielsen}. The purely classical case (${\cal A}$ 
abelian and coinciding with the center ${\cal Z}$) shows this is no more the case if the center is non trivial.    

\subsection{Entropy properties, relative entropy, and mutual information}
The above definition of entropy for a state on the algebra is the one which is completely intrinsic, i.e. depending only on the physical 
expectation values of the algebra operators and nothing else. Besides, it has a number of interesting properties which copy the ones for the 
usual case of reduced density matrices by partial tracing, and which will be useful in the analysis of the continuum limit, and relevant to 
some applications. In particular,  the proof of entropic c-theorems depends on strong subadditivity, and it is important to recognize this property 
is present in this more general algebraic setting. We list here the properties which we use in later discussion \cite{petz}.  

\bigskip

\noindent 1) If the global state is pure we have the {\sl symmetry property} 
\begin{equation}
S({\cal A})=S({\cal A}^\prime)\,.
\end{equation}
 This follows because the 
classical probabilities $p_k$ are shared by ${\cal A}$ and ${\cal A}^\prime$, and the density matrices $\rho_{{\cal A}_k{\cal A}_k^\prime}$ 
in (\ref{dua}) are pure. Hence $S(\rho_{{\cal A}_k})=S(\rho_{{\cal A}_k^\prime})$ for each sector $k$. 

\bigskip

\noindent 2) {\sl Strong sudadditivity}. This holds for three algebras in tensor product. This is the case for three 
algebras ${\cal A}$, ${\cal B}$, ${\cal C}$, which are mutually commuting and with trivial intersections (such as the algebras of spatially separated regions). We have 
\begin{equation}
S({\cal A}{\cal B})+S({\cal B}{\cal C})\ge S({\cal C})+ S({\cal A}{\cal B}{\cal C})\,.
\end{equation} 

\bigskip 

\noindent 3) The {\sl relative entropy} for two states in the same algebra can be defined in the usual way using the reduced density matrices, 
\begin{equation}
S(\rho^1_{\cal A}|\rho^0_{\cal A})=\textrm{tr} (\rho^1_{\cal A} \log \rho^1_{\cal A}- \rho^1_{\cal A} \log \rho^0_{\cal A})\,.\label{rela}
\end{equation}
Using the explicit form (\ref{explicit}) we have 
 \begin{equation}
S(\rho^1_{\cal A}|\rho^0_{\cal A})=\sum_k p_k^1 \log(p_k^1/p_k^0)+\sum_k p_k^1 S(\rho^1_{{\cal A}_k}|\rho^0_{{\cal A}_k})\,.\label{rerelala}
\end{equation}
The first term on the right hand side is the classical relative entropy of the two probability distributions, and the second is the average of the 
relative entropies of the quantum states on the different sectors.  The relative entropy is positive and monotonously increasing with inclusion of algebras,
\begin{equation}
S(\rho^1_{\cal A}|\rho^0_{\cal A})\le S(\rho^1_{\cal B}|\rho^0_{\cal B})\,\hspace{1cm} \textrm{for } {\cal A}\subseteq {\cal B}\,. 
\end{equation}

\bigskip

\noindent 4) The {\sl mutual information} for one state and two commuting algebras with intersection ${\bf 1}$ can be defined as a relative 
entropy of two states on the joint algebra ${\cal A}{\cal B}$, 
\begin{equation}
I({\cal A},{\cal B})=S(\rho_{{\cal A} {\cal B}}| \rho_{\cal A}\otimes \rho_{{\cal B}} )=S({\cal A})+S({\cal B})-S({\cal A} {\cal B})\,. \label{mutu}
\end{equation}
In the lattice, this is achieved for the algebras of two disjoint regions with no common element in the center. 

Taking into account that the center of ${\cal A}{\cal B}$ is the tensor product of the centers of each algebra,  
mutual information writes in terms of the states on each sector and the classical probabilities of the common eigensectors of the center
\begin{equation}
I({\cal A},{\cal B})=\sum_{k_A, k_B} p_{k_A, k_B} \log(p_{k_A, k_B}/(p_{k_A} p_{k_B}))+\sum_{k_A, k_B} p_{k_A ,k_B} S(\rho_{k_A, k_B}|\rho_{k_A}\otimes\rho_{k_B}) \,.\label{mi}
\end{equation}
The first term is again the classical mutual information for the probability distributions $p_{k_A}$ and $p_{k_B}$ on 
the centers of ${\cal A}$ and ${\cal B}$, given the joint probability distribution $p_{k_A, k_B}$ on the center of the full algebra.

Due to analogous properties of the relative entropy, mutual information is positive, and increasing with ${\cal A}$ and ${\cal B}$.

\section{Ambiguities in the correspondence of algebras and regions}

We have seen local algebras of gauge theories typically have a non trivial center. In this case, there is no interpretation as a tensor 
product structure of the Hilbert space. We have also shown this is no obstacle to compute a local entropy for the global state on the local algebra. In this 
section, we will explore more systematically some possible choices of algebras, as well as connect some particular algebra choice with previous 
constructions in the literature. We will also show there are always some choices of local algebras with trivial center. These are related to some special kind of gauge fixings.

\bigskip

\subsection{Two geometric choices}

\begin{figure}
\centering
\leavevmode
\epsfysize=5cm
\epsfbox{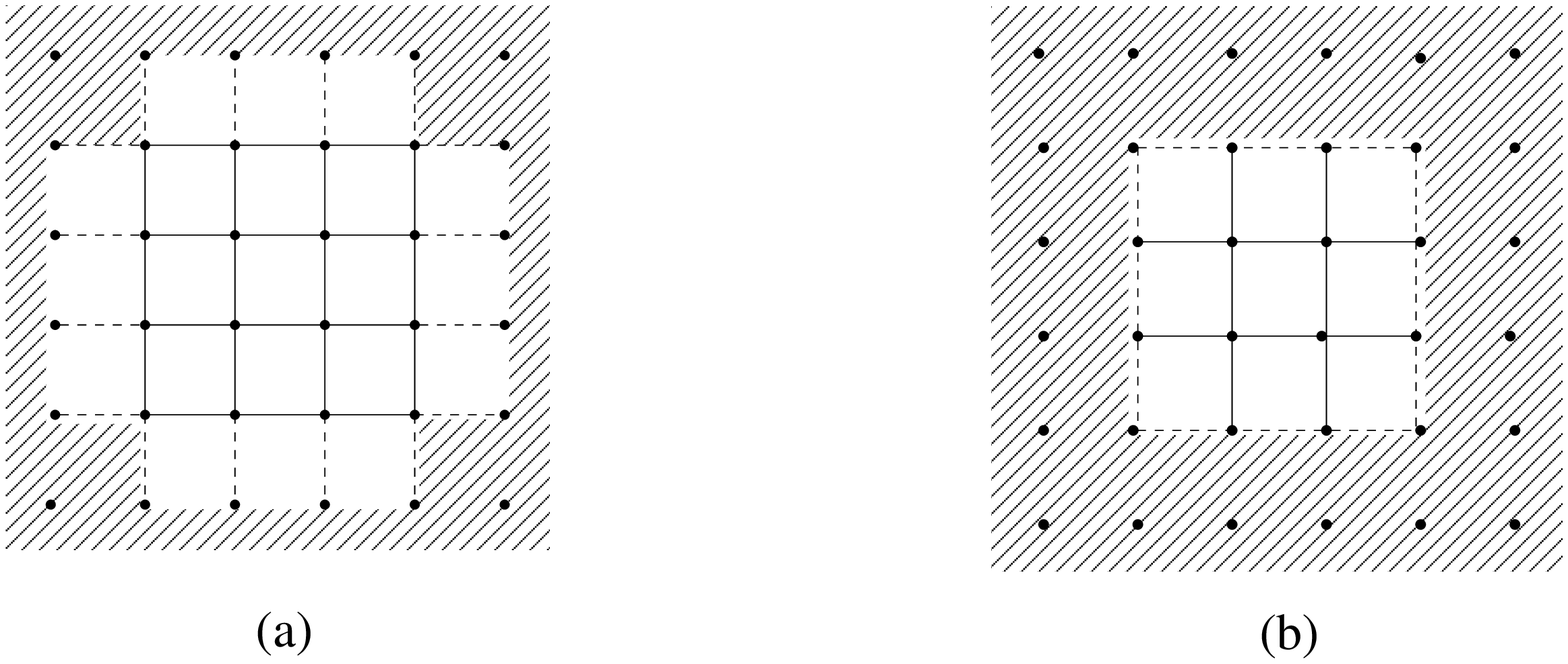}
\bigskip
\caption{(a) The algebra of the square with an electric center choice. The center is formed by the link operators shown with dashed lines. 
The commutant is represented by the shaded region, having the same center. (b) The square with a magnetic center choice. The center is formed 
by a single loop at the boundary in this two-dimensional example.}
\label{figu2}
\end{figure}

Let us consider some examples of assignations of algebras to regions. Our first choice is one (figure \ref{figu2}a) with a purely ``electric center''.  
With a given region $V$ (a set of links to be concrete) we take as ${\cal A}_V$ the algebra generated by all Wilson loops and link electric 
operators in $V$. This is the same choice of the previous section. In this case, the center is generated by all links operators of links with at 
least one vertex in $V$, and not forming part of any plaquette in $V$ (as shown in figure \ref{figu1}). With this choice, the region $\bar{V}$ is naturally the set of links separated by 
one link distance from $V$, and its algebra is ${\cal A}^\prime$, sharing the same center\footnote{Note we could have also defined the region $V$ 
for example as the square in figure \ref{figu2} plus all the links coming out of it, without modifying the content of the algebra, excepting at 
the corners. With this interpretation the intersection between $V$ and $\bar{V}$ is the set of all links in between, whose algebra is the common 
center.}. This is illustrated by the shaded region in figure \ref{figu2}a for a case of a two dimensional lattice. The full algebra generated 
by ${\cal A}\cup {\cal A}^\prime$ is not the algebra of all operators in Hilbert space because it does not contain the plaquettes at the boundary.

Hence, if we choose to keep all the operators that can be formed locally in $V$ there must be a center. Since the center is formed by products 
of link operators on the boundary we can follow a strategy of eliminating some of these link operators at the boundary in order to reduce the center. We have to keep 
all operators in the bulk of the region, otherwise at some point our algebra and the region are not anymore related to each other. Our second 
choice is then on the opposite extreme of the previous one, where we take out of the generating set of the algebra all link operators on the 
boundary. This purely ``magnetic center'' choice gives a center formed by all Wilson loops lying on the boundary (see figure \ref{figu2}b). Again 
the commutant ${\cal A}^\prime$ can be ascribed to the opposite region $\bar{V}$, with the same prescription.     

\subsection{The extended lattice construction}
 Buividovich and Polikarpov \cite{Polikarpov} used a specific construction for computing the entanglement entropy for gauge fields in the lattice and 
overcome the difficulties imposed by the constraints. This followed earlier work on topological models \cite{Wen,zanardi} and loop quantum 
gravity \cite{dongrav}. The construction was further developed by Donnelly in \cite{donnelly}. 

\begin{figure}
\centering
\leavevmode
\epsfysize=4cm
\epsfbox{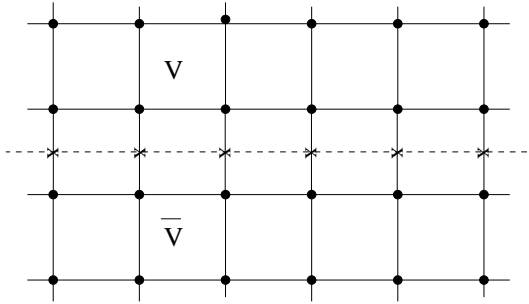}
\bigskip
\caption{The links cut by the boundary $\partial V$ (dashed line) are duplicated. The crosses are new vertices of the lattice, but 
no gauge invariance is required for them. }
\label{buido}
\end{figure}

The basic method consists in drawing a region $V$ on the lattice ${\cal L}$, such that the boundary $\partial V$ does not pass through any vertex. Consider a link $l_\partial$  which is cut in two by 
the boundary of $V$ (see figure \ref{buido}). We call the set of all these links $ {\cal L}_{\partial}$. In order to produce a tensor product of Hilbert spaces 
on each side of the boundary $\partial V$ a new vertex $a_{l_\partial}$ is introduced in the intersection of $\partial V$ and $l_\partial$, 
dividing $l_{\partial}$ in two links, $l_{\partial V}$ and $l_{\partial \bar{V}}$, one on each side of the boundary $\partial V$. Call these 
sets of links ${\cal L}_{\partial V}$ and ${\cal L}_{\partial \bar{V}}$, and the new ($V$-dependent) lattice ${\cal L}^\prime$. The Hilbert space  is 
then increased form the the original space ${\cal H}$ of gauge invariant function on the lattice links to the one ${\cal H}^\prime$ of 
functions on all links, including the new ones, which are gauge invariant with respect to the gauge transformations based on the old vertices 
but not necessarily gauge invariant with respect to gauge transformations based on the new vertices $a_\partial$ on the boundary.

 Since gauge transformations act now independently on each side of $\partial V$, the space ${\cal H}^\prime$ of functions on links which 
are gauge invariant in this last sense is the tensor product ${\cal H}^\prime_V\otimes {\cal H}_{\bar{V}}^\prime$ of the spaces of gauge 
invariant function on links on both sides of $\partial V$. A norm preserving linear mapping from gauge invariant functions in the original Hilbert 
space ${\cal H}$ to the new space ${\cal H}^\prime$ is given by
\be
\Psi^\prime(U_{l_1},...,U_{l_{\partial V}},U_{l_{\partial \bar{V}}},...,U_{l_N})\equiv\Psi(U_{l_1},...,U_{l_{\partial V}} . U_{l_{\partial \bar{V}}},...,U_{l_N})\,.\label{op}
\ee    
In (\ref{op}), the rule is understood for all original link $l_\partial$ on the boundary which became the two links $l_{\partial V}$ and $l_{\partial \bar{V}}$, where $l_\partial$, $l_{\partial V}$ and $l_{\partial \bar{V}}$ are all taken with the same orientation. In this way $U_{l_{\partial V}} . U_{l_{\partial \bar{V}}}$ replaces the original link variable $U_\partial$ in the wave function.

After this step, the reduced density matrix and entanglement entropy calculation follows the usual rule for tensor products,
\be
\rho_V=\textrm{tr}_{{\cal H}_{\bar{V}}} (\Psi \Psi^\dagger)\,.\label{up}
\ee
Let us call ${\cal L}_V$ to the links on $V$ in the new lattice, and ${\cal L}_V={\cal L}_{V^i}\cup {\cal L}_{\partial V}$, is a union of the links in the 
interior ${\cal L}_{V^i}$ of $V$ and the ones on the boundary. Analogously ${\cal L}_{\bar{V}}={\cal L}_{\bar{V}^i}\cup {\cal L}_{\partial \bar{V}}$.  We can 
write (\ref{up}) more explicitly, using  (\ref{op}), as
\be
\rho_V[U_{V^i},U_{\partial V},U_{V^i}^\prime,U_{\partial V}^\prime]=\int (\Pi_{l_{\bar{V}}\in {\cal L}_{\bar{V}}} dU_{l_{ \bar{V}}})\,\Psi[U_{V^i},U_{\bar{V}^i},U_{\partial V}U_{\partial \bar{V}} ]\Psi^*[U_{V^i}^\prime,U_{\bar{V}^i},U_{\partial V}^\prime U_{\partial \bar{V}}]\,.\label{episte}
\ee
For discrete gauge groups the integrals are replaced by sums. The entropy is then computed with the von Neumann formula.

The question which arises is in which sense this construction can be considered an entanglement entropy of the original model. 
The construction is clearly uniquely defined, but are there other possibilities? Or, in other words,  how it depends on external elements 
introduced to the original model? We now show this construction is equivalent to the electric center choice, and hence the entropy 
this method produces is the entropy of a possible choice for local gauge invariant algebra attached to $V$, though strictly speaking it is not an 
entanglement entropy in the sense of measuring entanglement in the original model.

To see this, note the new algebra of operators in $V$ is a full matrix algebra. This is because, contrary to the electric center choice, we 
have now new string operators which are analogous to the Wilson loops but they are not closed but open, having the two boundaries on the 
boundary vertices. Since these vertices do not produce gauge transformations, the open strings operators are gauge invariant. However, 
the expectation value of all original operators has not changed. For the operators generated in the interior $V^i$ this is evident from (\ref{op}). 
For the link operators in the boundary, this follows from 
\bea
&&\int (\Pi_{l\in {\cal L}^\prime} dU_l) \,\Psi^\prime(U_{l_1},...,U_{l_{\partial V}},U_{l_{\partial \bar{V}}},...,U_{l_N})^*\, \hat{L}^{l_{\partial V}}_{g}\, \Psi^\prime(U_{l_1},...,U_{l_{\partial V}},U_{l_{\partial \bar{V}}},...,U_{l_N})\nonumber\\
&&=\int (\Pi_{l\in {\cal L}^\prime} dU_l) \,\Psi^\prime(U_{l_1},...,U_{l_{\partial V}},U_{l_{\partial \bar{V}}},...,U_{l_N})^* \Psi^\prime(U_{l_1},...,g U_{l_{\partial V}},U_{l_{\partial \bar{V}}},...,U_{l_N})\nonumber\\
&&=\int (\Pi_{l\in {\cal L}^\prime} dU_l) \,\Psi(U_{l_1},...,U_{l_{\partial V}}U_{l_{\partial \bar{V}}},...,U_{l_N})^* \Psi(U_{l_1},...,g U_{l_{\partial V}}U_{l_{\partial \bar{V}}},...,U_{l_N})\nonumber\\
&&=\int (\Pi_{l\in {\cal L}} dU_l) \,\Psi(U_{l_1},...,U_{l_{\partial}},...,U_{l_N})^* \hat{L}_g^{l_\partial}\Psi(U_{l_1},..., U_{l_{\partial}},...,U_{l_N})\,,
\eea
where in the last line we have used normalization of the measure on the gauge group and a change of variables. 

Moreover, link operators $\hat{L}_g^{l_{\partial V}}$ in the boundary commute with the density matrix. From (\ref{episte})
\bea
\hat{L}_g^{l_{\partial V}}\rho_V[U_{V^i},U_{\partial V},U_{V^i}^\prime,U_{\partial V}^\prime]=\int (\Pi_{l\in {\cal L}_{\bar{V}}} dU_{l})
\,\Psi[U_{V^i},U_{\bar{V}^i},g U_{\partial V}U_{\partial \bar{V}} ]
\Psi^*[U_{V^i}^\prime,U_{\bar{V}^i},U_{\partial V}^\prime U_{\partial \bar{V}}]\nonumber
\\
= \int (\Pi_{l\in {\cal L}_{\bar{V}} dU_{l}}) \, \Psi[U_{V^i},U_{\bar{V}^i}, U_{\partial V}U_{\partial \bar{V}} ]\Psi^*[U_{V^i}^\prime, U_{\bar{V}^i},g^{-1} U_{\partial V}^\prime U_{\partial \bar{V}}] = \rho_V[U_{V^i},U_{\partial V},U_{V^i}^\prime,U_{\partial V}^\prime]\hat{L}_g^{l_{\partial V}}\,.
\label{episte1} 
\eea

Hence, even if the algebra is increased with respect to the case of the electric center, the density matrix is block diagonal in the basis 
which diagonalizes the link operators in the boundary, and produces the same expectation values as the original state on the algebra of the 
electric center choice. The entropy (\ref{er}) is then the same, proving our assertion that the Buividovich-Polikarpov 
method is equivalent to the electric center choice.
\subsection{Local algebras with trivial center and entanglement entropy}
There are many possibilities for choosing boundary details of the algebra, and these can be ordered by inclusion: by peeling off  
the boundary link operators we convert the electric center to a magnetic one, and taking out the boundary loops we come again to the electric 
center, but in a reduced region. Interestingly, in taking out of the algebra link operators at the boundary in passing from the 
electric to the magnetic center, at some point we could reach a trivial center case, with balanced number of magnetic and electric operators. We now 
study how this can be achieved.  

To analyze the constraints in geometric terms, let us first think in terms of an abelian gauge field in the continuum. The algebraic description of the gauge theory is in terms of gauge invariant 
operators $\vec{E}$ and $\vec{B}$ (in $d=3$). These are not all independent because they have to obey the time independent electric and 
magnetic Gauss laws, $\vec{\nabla}.\vec{E}=0$ and  $\vec{\nabla}.\vec{B}=0$. In general spacial dimension $d$ these write 
\be 
\nabla_i F^{0 i}=0\,, \hspace{2cm}   \partial_{i_1}  F_{i_2 i_3}+\partial_{i_2}  F_{i_3 i_1}+\partial_{i_3}  F_{i_1 i_2}=0\,.\label{constri}
\ee 

\begin{figure}
\centering
\leavevmode
\epsfysize=5cm
\epsfbox{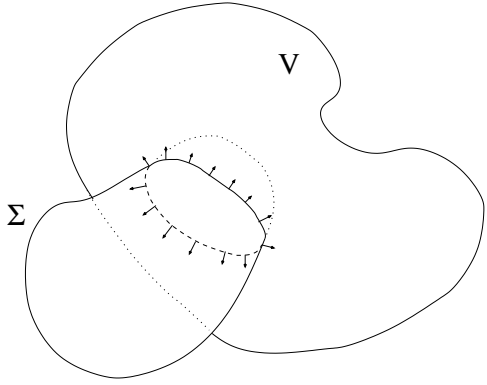}
\bigskip
\caption{A surface $\Sigma$ intersecting the region $V$. If the link operators on $\partial V$ coming out of $\Sigma$ are in the algebra ${\cal A}_V$, this algebra has a non trivial center as a result of the Gauss law constraint.}
\label{gauss}
\end{figure}

Integrating the first of equations (\ref{constri}) on a $d$-dimensional volume $W$ bounded by a closed $(d-1)$-dimensional surface $\Sigma$, we have
\begin{equation}
\int_{W} d^{d}x\,\vec{\nabla}.\vec{E}=\int_\Sigma d\sigma\, \vec{\eta}.\vec{E}=0\,.\label{con1}
\end{equation}
Hence, for any surface $\Sigma$ intersecting our region of interest $V$ (see figure \ref{gauss}), the operator 
\begin{equation}
\int_{\Sigma \cap V} d\sigma \, \vec{\eta}.\vec{E}=-\int_{\Sigma \cap \bar{V}} d\sigma \, \vec{\eta}.\vec{E}\,,\label{mis}
\end{equation}
is a potential element of the center of ${\cal A}_V$. This is because (\ref{mis}) is equal to some operator formed by the electric field  
outside $V$, and consequently commutes with the rest of the algebra. Of course, if this operator belongs to the center or not can be controlled 
by details on how the  algebra is chosen on the surface of $V$.  

In the same way, integrating the second equation of (\ref{constri}) in a three dimensional volume we get for any closed two-dimensional surface $\Sigma$ the integral $\int d\sigma_{ij} F^{ij}=0$. 
This gives
\be 
\int_{\Sigma\cap V} d \sigma_{ij}\,F^{ij}=-\int_{\Sigma\cap \bar{V}} d \sigma_{ij}\,F^{ij}=\oint_{\partial V\cap \Sigma} A_\mu.dx^\mu\,.\label{con2}
\ee 
Hence, in the lattice we can form an element of the center in this way if the corresponding Wilson loop on the boundary of $V$ belongs to the algebra,  and the algebra does not 
contain any operator not commuting with it. 

The lattice versions of (\ref{con1}) and (\ref{con2}) are easily constructed. Eq. (\ref{con1}) follows by multiplying the constraint 
equation $\Pi_{a\in W} \hat{T}_{g_a}=1$ for all vertices in a region $W$. In this equation expressed in terms of link operators $\hat{L}$ all 
link operators between two vertices in $W$ cancel exactly since they appear with opposite directions. Then, this is equivalent to 
\be
\Pi_{l \in \partial W_{\textrm{out}}} \hat{L}^l_g=1\,,
\ee
where $\partial W_{\textrm{out}}$ is the set of links with one and only one vertex on $W$, pointing outwards. The constraint implies the 
link operators for a fixed group element $g$ attached on one side of (and non included in) any closed $(d-1)$-dimensional surface have product equal to $1$. 

Analogously, given an oriented closed two-dimensional surface on the lattice, the product of the oriented plaquettes (with the same representation)
equal to one, because all links appear once with each orientation. This gives the lattice version of (\ref{con2}). 

Now we can establish the conditions for the algebra of a region $V$ to have trivial center. 
Our strategy is to start with all link and Wilson loop operators  
that can be drawn on $V$, and eliminate some link operators on the surface of $V$. 
Let us call $\partial V^+$ to the set of links on the 
boundary $\partial V$ which are represented by link operators in the algebra and $\partial V^-$ to the links on the boundary whose link operators 
do not belong to the algebra. The two different constraint equations imply two conditions for the distribution of link operators which we allow on the surface:

\bigskip

\begin{figure}
\centering
\leavevmode
\epsfysize=4cm
\epsfbox{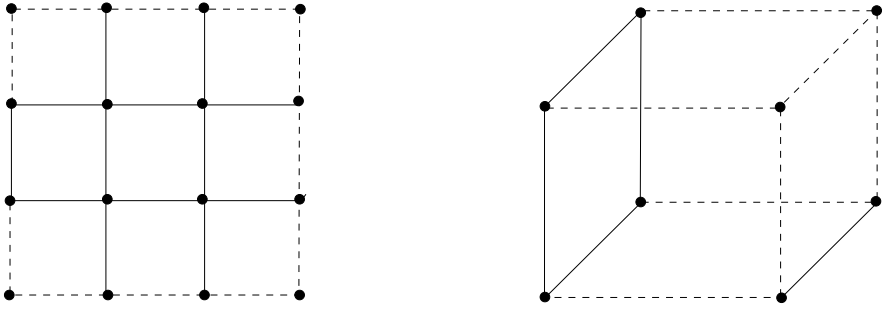}
\bigskip
\caption{Left panel: Choice of an algebra with trivial center for $d=2$. A maximal tree on the boundary is shown with a dashed line. Only one link operator on the boundary of $V$ is included in ${\cal A}_V$. Two links on the boundary lead to a non trivial center through the electric Gauss law. In this example 
the total number of electric and magnetic degree of freedom equals $9$: $9$ plaquettes, and $13$ link operators with $4$ constraint equations.     
Right panel: 
tree of links on the surface (dashed line). The number of electric and magnetic degrees of freedom equals $5$ in this example (there are $6$ plaquettes
but the product of all plaquettes is $1$).}
\label{figu4}
\end{figure}

\noindent a) In order that electric constraint cannot be used to produce an element on the center we have to avoid having a set of link operators 
on the boundary which are attached and external to one side of a closed $(d-2)$-dimensional surface on the $(d-1)$-dimensional boundary $\partial V$. 
This is equivalent to say that $\partial V^-$ has to be taken as a connected set (i.e. it cannot be divided in two by the set of links which are chosen 
to belong to the algebra). We must add to this the trivial case: The set of link operators on the boundary which belong to the algebra should not 
contain all the links attached to a single point in $\partial V$, or equivalently, $\partial V^-$ must pass to every vertex on $\partial V$.     

\bigskip

\noindent b) Magnetic constraints imply in order to have no center the set $\partial V^-$ must not contain any closed path. Otherwise its Wilson 
loop operator belongs to the center. Hence, this condition is that $\partial V^-$ is a tree.

\bigskip

Conditions (a) and (b)  combined give the following prescription: In order to produce an algebra ${\cal A}_V$ with a trivial 
center we can take all operators which can be drawn on $V$ excepting for the link operators corresponding to a {\sl maximal tree} of links drawn on the surface of $V$. 

It is evident that such maximal trees always exist, and we can always choose a local algebra with trivial center. Figure \ref{figu4} shows 
two examples in $d=2$ and $d=3$. The form of these maximal trees are highly arbitrary. The fact that the center 
is trivial, and the entropy is an entanglement entropy, do not eliminate the ambiguities. Further, in contrast to the universal 
geometric prescriptions of electric and magnetic centers of the previous discussion, the the particular choice of a maximal tree on the surface  unavoidably breaks lattice symmetries.    

With this choice the commutant algebra is also a full matrix algebra, and has the same local structure with respect to 
some region $\bar{V}$, which now depends on the maximal tree. More explicitly, ${\cal A}(V)^\prime$ can be thought as arising from the region $\bar{V}$
 formed by all links $l$ on the lattice such that $\hat{L}_g^l\notin {\cal A}(V)$. The algebra ${\cal A}(V)^\prime$ is obtained from all Wilson loops in this $\bar{V}$ and all link operators which are not in the maximal tree. The complementary regions $V$ and $\bar{V}$ share the same boundary maximal tree.  

\subsection{Algebra choice by gauge fixing}

The maximal tree establishes a natural connection with gauge fixing in the lattice. In this section we show the above choices of 
algebras with trivial center are in correspondence to some special gauge fixings. 

Suppose we want to fix the gauge in the lattice. 
A single link variable $U_{(ab)}$ can be fixed to $1$ by sacrificing the gauge freedom of one of the vertices $a$ or $b$. In this way we can 
start fixing different link variables to $1$. However, we cannot fix to $1$ all links in a closed path, since that would contradict the gauge invariance 
of the product of link variables along the path. Hence, the links that can be fixed have to form a tree. In fact, we can always fix a maximal 
tree: If there is a link which has not been fixed, but which does not closes a loop to the tree of already fixed links, then it must be that 
some of its two end-points does not belong to the tree, and whose corresponding gauge transformation has not been used. Then, we can use this gauge 
transformation to fix the new link.

In this way we can map a gauge theory into a theory where local gauge invariance is absent, and physical variables $U_l\in G$ are attached to the 
links complementary to the maximal tree of fixed links. Let us call $T$ to the maximal tree and $\bar{T}$ to the complementary set of links. 
Note that each of the physical $U_l\in \bar{T}$ actually expresses the value of the product of link variables on a closed loop that is equal 
to $U_l$ times some variables along the maximal tree, which are now set to $1$. Also, the gauge invariant wave functions are just the 
ordinary functions on $\bar{T}$.

Then, since the variables in $\bar{T}$ describe all the independent loop variables, the algebra of Wilson loop operators is generated 
by the coordinate operators $\hat{U}_l^r$, $l\in \bar{T}$, with
\be
\hat{U}_l^{r} \Psi[U_1,..., U_l,...,U_N]=U_l^{r} \Psi[U_1,..., U_l,...,U_N]\,,
\ee     
where $N$ is the total number of links on $\bar{T}$.

In this gauge fixed representation is straightforward to define tensor products and entanglement entropy. We just have to select some 
subset $V\subseteq \bar{T}$ of non-fixed links and its complement $\bar{V}$. The tensor product 
decomposition ${\cal H}={\cal H}_V\otimes {\cal H}_{\bar V}$ of the arbitrary functions on $\bar{T}$ directly gives an entanglement entropy for $V$. 

\begin{figure}
\centering
\leavevmode
\epsfysize=5cm
\epsfbox{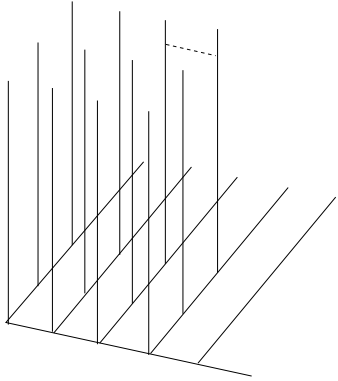}
\bigskip
\caption{Particular choice of maximal tree for three dimensions. The dashed link is 
associated to a large loop extended in the vertical direction. }
\label{axial}
\end{figure}

The pitfall in this construction is that while this clearly gives an entanglement entropy for some decomposition of the global 
Hilbert space, in general, there is no relation between this decomposition and entanglement entropy of a spacial region. The degree 
of freedom labeled by the gauge invariant links in $\bar{T}$ can indeed be highly non local with respect to the usual localization of 
operators. To see this, consider the axial gauge fixing given by the maximal tree of figure \ref{axial}. Any link variable $U_l$ for $l\in \bar{T}$ 
describes the holonomy corresponding to a potentially very large loop extended in one direction of space. The entropy of some set of 
links in $V\cap \bar{T}$, with $V$ some region of the space, will not describe the actual entropy in $V$ but something very different, 
which is sensible for example to details of the state very far away from $V$.   

Hence, if we want to retain a meaning of localization, we have to make sure in closing some path of the maximal tree with a link 
in $\bar{T}\cap V$ this loop is also contained in $V$. And this is exactly what the prescription of the previous subsection manages 
to do. By choosing a maximal tree on the surface of $V$ any link on $\bar{T}\cap \partial V$ necessarily closes a loop on $\partial V$ 
and does not go far away. If we extend the maximal tree on $\partial V$ to all the space (what can always be done), closing the tree 
with a new link $l$ inside of $V$, the corresponding loop cannot pass through the boundary of $V$. To see this suppose that this loop 
starting in $l\in V$ actually passes though $\partial V$ and closes in $\bar{V}$. Then it passes through two vertices on $\partial V$ 
which are joined by a part of $T$ outside $V$. But this is not possible, since any two points in $\partial V$ already determine a path 
in $T$ on the surface $\partial V$, and hence the previous assumption would imply the existence of a closed loop in $T$. 

Therefore, a maximal tree on the surface $\partial V$ effectively cuts the degree of freedom in two, the ones inside and outside $V$. 
The relation of the unfixed links in $\bar{T}$ to actual localized operators on the lattice can still be very fuzzy, but the mapping 
is non local mixing operators inside and outside $V$ among themselves, and this is the only thing that is needed for a local entropy in $V$. 
In fact, as we have seen, the rest of the tree extending the one in $\partial V$ is not necessary, and does not change the entropy. We can just make a partial gauge fixing along the boundary to 
get an entanglement entropy.

Generators for the rest of the gauge invariant operator algebras can be chosen naturally as the link operators $\hat{L}_g^l$ with $l\in \bar{T}$. 
Then the fact that the maximal tree in $\partial V$ can be extended to a maximal tree inside $V$ explains that the counting of electric and magnetic degrees of freedom matches inside the algebra of $V$ defined in this way. This  gives place to a full matrix algebra. For 
this counting see for example \cite{bishop}.

\section{How ambiguous is the entropy?}
We now want to have an idea on how much can the entropy change with a change of prescription at the boundary. We analyze first the 
simpler case of a free scalar field which is very instructive in this respect. Then we analyze a topological $Z_2$ model which has been 
previously discussed in the literature. 

\subsection{Scalar field in the lattice}

Suppose that in analogy with the case of the gauge field we change the natural prescription to compute the entropy in a region $V$ 
for a scalar, and take as the algebra of $V$ all field and momentum operators of the interior points of $V$ but we only take the field 
operators, and not the momentum operators (or viceversa), for the points on a subset $A$ of the boundary. This leads to a center generated by all field operators $\hat{\phi}_a$ with $a\in A\subseteq  \partial V$. 

A simultaneous diagonalization of the center is achieved in the basis of eigenvectors of the 
variables $\hat{\phi}_a|\{\phi\}\rangle=\phi_a |\{\phi\}\rangle$. The probability distribution for the different 
basis vectors in the vacuum state has to be a Gaussian distribution, since it must reproduce the expectation value of product 
of the fields, which satisfies the Wick theorem. Then, we have 
\begin{equation}
p(\{\phi\}_A)=\sqrt{\det(M_A/(2\pi))} \,\,e^{- \frac{1}{2}\phi_a M_A^{a b}\phi_b}\,,\label{ffi}
\end{equation}
where the matrix $M_A$ is adjusted such that the two point function of the field is reproduced by the probability distribution
\be
X^A_{ab}\equiv\langle \hat{\phi}_a \hat{\phi}_b\rangle_A=\int (\Pi_{a\in A} d\phi_a) \, p(\{\phi\}_A)\, \phi_a \phi_b=(M_A^{-1})_{a b}\,.\label{ffi1}
\ee

An interesting point here is that (\ref{ffi}) is a probability distribution over a continuous variable, or a set of continuous variables.
It is known there is no unambiguous entropy for continuous variable distributions. One can convince oneself that the would be classical entropy formula 
\be
H=-\int (\Pi_{a\in A} d\phi_a) \, p(\{\phi\}_A) \log(p(\{\phi\}_A))=  \textrm{tr} \left(\frac{1}{2}-\log\left(\frac{M}{2 \pi}\right)\right)\,\label{34}
\ee
is ambiguous because the probability $p(\{\phi\}_A)$ is now a probability density, and it has some dimensions which are not 
compensated in the logarithm. In other words, this formula is not invariant under redefinitions of the field variable.

Another interesting aspect of (\ref{34}) is the following. For sufficiently large $M$ the field becomes highly concentrated, 
with large $p(\{\phi\}_A)$ for some part of the field space. In consequence the logarithm of the entropy density is not negative 
any more and the resulting entropy is negative. This also points to the ambiguous nature of the classical entropy for continuum 
variables, but it reminds us of some puzzling negative entropy terms for continuous gauge groups found in the literature \cite{Kabat}.

If in order to make sense of (\ref{34}) the variable $\phi_a$ is discretized, the probability for each value of $\phi$ goes 
as $p\sim N^{-1}$ with the number of discrete values of $\phi$, to keep the distribution normalized. Then the probability 
distribution has an infinite entropy limit $H\sim \log(N)$ as $N\rightarrow \infty$. This infinite is different from the 
divergences of the entropy in the continuum limit of the lattice, and is there also for a single harmonic oscillator.

 This example shows in the discrete case the classical part of the entropy can be very large if the center has a large number 
of sectors. This means also that the numerical values of the entropy will typically change by large amounts for different 
prescriptions varying details of the algebras on the boundary. In particular, if the gauge group is continuous and the center 
contains at least one Wilson loop operator we expect the entropy to be ill-definite, and depending of our choice it can be infinite 
or even negative. If the gauge group is continuous and the gauge group is non compact, the same happens if the center contains 
any link operators.  These cases with center formed by continuous variables is the worst case for the entropy ambiguities, where minor modification of 
the algebra leads to big entropy changes.

\subsection{Topological $Z_2$ model}

Consider a $Z_2$ gauge model on a square lattice in $d=2$. For each link there is a variable $z_l$ taking values in $\{1,-1\}$. We take the wave 
function of a topological model, which has been studied previously in the literature \cite{Polikarpov,donnelly,vidal}. The state can be mapped  
to the toric code spin model, though the complete Hilbert space is different \cite{kitaev97}. The wave function is
\be
\Psi[z_l]=K \,\sum_{\Gamma}\,\Pi_{l\in \Gamma}\,\, z_l\,, 
\ee
where the sum is over all closed paths $\Gamma$, containing any number of closed loops, intersecting or not, and $K$ is a normalization constant.
More precisely, $\Gamma$ can be any set of links such that for any vertex there is an even number of links in $\Gamma$ attached to the vertex. 
It is evident this wave function is gauge invariant because it is a superposition of Wilson loops operators acting on the function $\Psi_0[z_l]\equiv 1$,
\be
\Psi[z_l]=K \,\sum_{\Gamma}\,W_\Gamma\,\, \Psi_0[z_l]\,. 
\ee
This wave function can be also defined as the unique common eigenvector of all Wilson loop operators with eigenvalue $1$,
\be
W_\Gamma \Psi[z_l]=\Psi[z_l]\,.
\ee
This follows from the fact the Wilson loops operators form a group, where $W_{\Gamma_1}W_{\Gamma_2}=W_{\Gamma_1 + \Gamma_2}$, and $\Gamma_1+\Gamma_2$ 
is the set of links which are in $\Gamma_1$ or in $\Gamma_2$ but not in both of them. The inverse is the same Wilson loop, because in this $Z_2$ 
model $W_\Gamma^2=1$, and the two orientations of a loop are equivalent. Hence, following \cite{zanardi}, we have
 \be
 W_\Gamma \Psi[z_l]=K \,\sum_{\Gamma^\prime}\,W_\Gamma W_{\Gamma^\prime}\,\, \Psi_0[z_l]=K \,\sum_{\Gamma^\prime}\,W_{\Gamma^\prime}\,\, \Psi_0[z_l]=\Psi[z_l]\,.\label{52}
 \ee

Link operators $\hat{L}^l_{-1}$ are identified with the Pauli matrices $\sigma_x^l$ acting on the basis diagonalizing the variables $z_l$. Note that $W_\Gamma \sigma_x^l=\sigma_x^l W_\Gamma$ if $l\notin \Gamma$, and $W_\Gamma \sigma_x^l=-\sigma_x^l W_\Gamma$ if $l\in \Gamma$. Hence 
we can write any product on the generators of the algebra (Wilson loops and link operators) as a product of link operators on the left and a 
Wilson loop on the right. The expectation value in $|\Psi\rangle$ is equivalent to the expectation value of the product of link operators alone. 
Now, the expectation value $\langle\Psi|\sigma_{l_1}\hdots\sigma_{l_k}|\Psi\rangle $ 
is either $1$ if $\sigma_{l_1}\hdots\sigma_{l_k}=1$, that is, if this is the expression of a constraint, or otherwise it is zero. To see this, 
note a constraint is a product of link operators for links with one vertex on a closed line. Then, a loop can remain on each side of this 
closed line or enter and come out of it, crossing the constraint surface an even number of times. If the product of link operators is not a 
constraint, there is a Wilson loop $W_\Gamma$ which passes through an odd number of the links $l_1$,...,$l_k$, and then
\be
\langle\Psi|\sigma_{l_1}\hdots\sigma_{l_k}|\Psi\rangle=\langle\Psi|\sigma_{l_1}\hdots\sigma_{l_k}W_\Gamma|\Psi\rangle=-\langle\Psi|W_\Gamma\sigma_{l_1}\hdots\sigma_{l_k}|\Psi\rangle=0\,.
\ee

Therefore if ${\cal O}_1$ and ${\cal O}_2$ are operators formed with linear combinations of products of some link operators and Wilson loops, 
 such that it is not possible to form any new constraint equation using the link operators used in ${\cal O}_1$ with the ones used for ${\cal O}_2$, we have 
\be
\langle \Psi|{\cal O}_1{\cal O}_2|\Psi\rangle=\langle \Psi|{\cal O}_1|\Psi\rangle\langle\Psi|{\cal O}_2|\Psi\rangle\,.\label{reli}
\ee  
This is because in the decomposition of the operators in sums of products of generators only the term with no link operators will survive 
the expectation value, and for this the relation (\ref{reli}) is a direct consequence of (\ref{52}). Relation (\ref{reli}) means there are 
no correlations in this model except the ones introduced by the constraint equations. In particular, for separated regions there are no 
correlations at all. This is why the model is topological.

Let us start computing the entropy of a region with the electric center choice. This has been considered previously in the literature, 
and it is equivalent to the extended lattice approach. We will reason algebraically here.  Suppose we have an algebra with a center 
generated by a set of $N$ independent link operators $\sigma_x^l$, or products of link operators, as $\sigma_x^{(6\, 9)} \sigma_x^{(6\, 10)}$ in 
figure (\ref{figu1}). That this set is independent means we have eliminated all operators which can be obtained from other by using a 
constraint equation. All these $N$ generators have eigenvalue $\pm 1$, and there are $2^N$ sectors labeled by 
eigenvalues $\lambda=\{\pm 1,\hdots,\pm 1\}$ for all the $N$ generators in the center. Because the expectation value for any 
multiple product of these operators always vanish, the only possibility is that the probability for all these sectors is the same $p_\lambda=2^{-N}$. 
Hence, the classical entropy is $H=N \log(2)$. It is easy to see for the electric center the number of independent generators in the 
center is $L-n_\partial$, with $L$ the number of links on the boundary, and $n_\partial$ the number of boundaries (multiple boundaries 
exist for non simply connected regions or regions with several connected components). This is because there is one independent constraint 
for each boundary. Hence $H=(L-n_\partial)\log(2)$.     

Now we have to evaluate the quantum entropy of the algebra once the simultaneous eigenvalues of the operators in the center have been 
fixed to $\lambda$. The algebra ${\cal A}_\lambda$ restricted to the sector $\lambda$ is formed by 
operators ${\cal O}_V^\lambda=P_\lambda {\cal O}_V P_\lambda$, where $P_\lambda$ is the projector to the sector $\lambda$. 
This is the product of projectors of the form
\be
\frac{1}{2}(1\pm \sigma_x^l)\,,
\ee  
where $\sigma_x^l$ belongs to the center.  
According to (\ref{explicit}) we have
\be
p_\lambda \textrm{tr}(\rho_V^\lambda {\cal O}_V^\lambda)=\langle \Psi|{\cal O}_V^\lambda|\Psi \rangle\,.\label{rry}
\ee
A complete set of commuting operators for ${\cal A}_\lambda$ is given by the projected Wilson loops in $V$. These Wilson loops 
commute with the center and have expectation values
\be
\langle \Psi|W_\Gamma^\lambda|\Psi \rangle=\langle \Psi|P_\lambda W_\Gamma P_\lambda|\Psi \rangle=\langle \Psi| W_\Gamma P_\lambda|\Psi \rangle=\langle \Psi|P_\lambda|\Psi \rangle=2^{-N}\,.
\ee 
Hence, from (\ref{rry}) 
\be
\textrm{tr}(\rho_V^\lambda W_\Gamma^\lambda)=1
\ee
for all Wilson loops in $V$. This implies in the basis which diagonalizes Wilson loops the state $\rho_V^\lambda$ 
has a diagonal with only one $1$ and all other entries equal to $0$. Hence, this is a pure density matrix with zero 
entropy\footnote{It cannot have off diagonal non-zero entries because it must be positive definite. Equivalently, if 
it has non-diagonal non-zero entries it is $\textrm{tr}((\rho_V^\lambda)^2)>1$, what is not possible.}. We conclude the 
quantum part of the formula (\ref{er}) is zero and the entropy with electric center writes
\be
S_E(V)=(L_V-n_\partial) \log (2)\,.\label{ele}
\ee 

For the case of the magnetic center, the center is formed by one Wilson loop for each independent boundary of $V$. The 
expectation values of all these loops is always $1$, and the probability of having the $-1$ eigenvalues is always zero. There 
is only the common eigenstate with eigenvalue $1$ for all loops which has probability one, and all other eigenstates have 
probability $0$. The classical entropy of the center vanishes. 

Then, there is only one sector with non-zero probability. 
Inside this sector we apply the same reasoning as above. The expectation values for all loops in this sector are unchanged with 
respect to the global state. This leads to a pure density matrix, and the entropy for a magnetic center vanishes,
\be
S_M(V)=0\,.
\ee 

For the case of trivial center we have of course that expectation values are equal to the ones given by the global state. These 
are again $1$ for each loop. It immediately follows the entanglement entropy is zero since the local density matrix is pure,
\be
S_{\textrm{ent}}(V)=0\,.
\ee 

Therefore, we see in this topological model the entropy has large variations, ranging from zero to an area law (\ref{ele}), according to the 
prescription for selecting the algebra. This is in part to be expected since there are no bulk correlations giving place to entanglement entropy but only 
entropy generated by our boundary prescription. In particular, our prescription for entanglement entropy gives zero entropy.

\subsubsection{Topological entanglement entropy}

\begin{figure}
\centering
\leavevmode
\epsfysize=4cm
\epsfbox{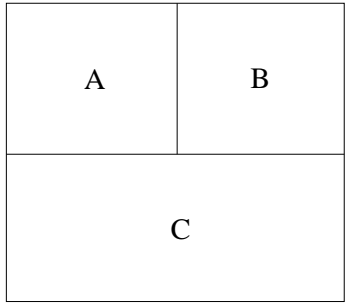}
\bigskip
\caption{Kitaev-Preskill tripartite geometric arrangement for calculating topological entropy.}
\label{topologica}
\end{figure}

The result (\ref{ele}) has been used to compute the topological entanglement entropy for this model. Topological entanglement 
entropy is a quantity which measures topological order for gapped systems. According to Kitaev and Preskill \cite{kitaev} this is 
defined through the combination of entropies for different regions as shown in figure \ref{topologica},
\be
S_{\textrm{topo}}=-\gamma_{\textrm{topo}}=S(A)+S(B)+S(C)-S(AB)-S(AC)-S(BC)+S(ABC)\,.\label{topo}
\ee
There is also an equivalent definition by Levin and Wen which differs from this one on the geometry \cite{Wen}.  

If we use equation (\ref{ele}) for the entropy with electric center we obtain 
\be
\gamma_{\textrm{topo}}=\log(2)\,, \label{pppo}
\ee
which is in accordance with the general formula for $\gamma_{\textrm{topo}}$ in terms of the quantum dimensions of the topological theory \cite{Wen}.
The area terms cancel in the combination (\ref{topo}).

However, if we use the entanglement entropy or the magnetic center entropy for the different regions we get 
\be
\gamma_{\textrm{topo}}=0\,. 
\ee
The topological entropy is then dependent on how the algebras are defined. In this example the presence 
of a center, and hence the fact that the entropy is not an entanglement entropy, is fundamental to have a non-zero result.

Perhaps some additional algebraic compatibility conditions are necessary to make it well defined. However, we have essayed some 
possibilities without success. For example, one could ask the algebras of unions such as $AB$ to be the generated algebras of $A$ and $B$, 
and also that the algebras of different regions are included in the commutants, for example ${\cal A}_B\subset {\cal A}_A^\prime$, etc.. 
This seems a straightforward requirement from the algebraic point of view. With these idea in mind the naive calculation with the electric 
center is changed to take into account for example the center in the algebra $AB$ formed by the common links operators at the boundary of $A$ and $B$ inside $AB$. It can be shown the final result is unchanged with this prescription. However, these requirements do 
not change the magnetic center calculation either and it still gives $\gamma_{\textrm{topo}}=0$.
 
The combination (\ref{topo}) is devised such that all local terms on the boundary (area terms, vertex terms), which might depend on short 
length physics and are not due to long range topological order, cancel. Then $\gamma_{\textrm{topo}}$ does not change with variations of the 
shape of the regions (for gapped models). However, the proof of topological invariance of $\gamma_{\textrm{topo}}$ and hence its physical content, 
requires some properties of the entropies which are not clearly displayed by different prescriptions.  First, it is important the details about how 
the boundary is locally defined in some trait are the same for any region in the combination (\ref{topo}) having this same trait at the boundary. 
For example, this is the case of the two intervals and the vertex on the boundary of $A$ which are also boundary of $AB$, $AC$ and $ABC$. If the local trait 
is shared in complementary way by two regions, the details of the algebras have to be as in the commutant algebras, such that the local 
contribution is the same. This is the case for example of the shared side of $A$ and $B$, or the   
 angle of $A$ on the point where $A$, $B$ and $C$ meet. This angle must have the same contribution as the complementary angle in $B\cup C$. 
 This seems to require the boundary details between $B$ and $C$ disappear in $BC$, and this is not the case for both the electric and magnetic center 
 choices if the algebra of $BC$ is the one generated by the algebras of $B$ and $C$. Even using the trivial center prescription, it seems in the present model there is no choice of algebras for all regions such that the algebras for the unions contain no bulk holes, in the sense of missing local generators. This is not the case if this topological vacuum state is interpreted as a state in a spin model as in \cite{Wen}. The spin model calculation matches the one here with the electric center, and may be the reason it gives the right topological entropy.   

\section{The continuum limit}

The entropy depends on the choice of algebra for $V$, and we have seen it can widely vary with this choice. Other quantities such as 
the mutual information $I(V,W)$ between two disjoint regions and the relative entropy $S(\rho^1_V|\rho^0_V)$ between two states in $V$ 
also depend on the algebra choice, and suffer from ambiguities in the lattice. However, these ambiguities are milder than the ones for 
the entropy. The important property of these quantities of being increasing with inclusion of algebras implies for example that 
the relative entropy for two states on $V$ with the magnetic center choice is bounded above by the one computed with the electric 
center choice, and bounded below by the relative entropy computed with the electric choice but for a region $V^\prime$ which is 
obtained from $V$ by taking out the boundary links. Likewise, all choices with trivial center are bounded by the same quantities. In 
this sense, mutual information and relative entropy vary smoothly in the lattice. 

As relative entropy quantities are expected to be finite in the continuum limit, this immediately implies 
that all choices we have made for the algebra ${\cal A}_V$ give the same relative 
entropy (and mutual information) in the continuum limit. This is because for all possible choices of boundary details in the algebra relative entropy  is bounded above and below by some quantities whose difference vanish with the cutoff. For example, if $V$ is characterized by a scale $R$, the relative entropy in the continuum is a smooth function $f(R)$, and any change of prescription involving a change of algebra in a region of depth $\epsilon$ on the surface produces changes in the relative entropy bounded above by $|f(R)^\prime \epsilon|$. This vanishes with $\epsilon$. 

There are good reasons to expect relative entropy quantities are finite in the continuum limit. This can be readily seen in the 
formulas (\ref{rela}) and (\ref{mutu}). These involve some subtraction of quantities computed with two density matrices in the same 
region. In the continuum limit the entropy has non universal divergences of ultraviolet origin, but these must be local on the 
boundary of the region, and thus get subtracted in the combinations (\ref{rela}) and (\ref{mutu}). 

Hence, the different possible choices of the boundary details of the algebras produce changes in the entropy which are however buried 
in the non universal terms local on the boundary in the continuum limit. That is, ambiguities in the continuum limit are of the 
same nature as the ones for other fields, and consist of terms local in the boundary of the region. 
Entropy is ill defined in the continuum because of 
these terms, both for gauge or other fields. 
Mutual information can be used to provide a universal geometrically regularized entropy in the form 
\be
S_\epsilon(V)=\frac{1}{2}\,\, I(V-\epsilon/2,\bar{V}+\epsilon/2)\,,\label{muyu}
\ee
 where $\epsilon$ is a short distance cutoff and $V- \epsilon/2$ is the region formed by $V$ by going inwards a distance $\epsilon/2$ from the boundary, and 
analogously $\bar{V}+\epsilon/2$ is formed from $\bar{V}$ going outwards a distance $\epsilon/2$. Note that for a finite system the limit 
$\epsilon\rightarrow 0$ gives the entropy since $S(V)=S(\bar{V})$ and $S(V\bar{V})=0$ for the total system in a pure state. The prescription (\ref{muyu}) is analogous to the framing regularization for Wilson loops. 
This can be used for example 
to give a proof to the c-theorem in $d=2$ \cite{mutualc}.

What is the significance of the classical Shannon term in the continuum limit? 

To give an answer to this question, let us consider the simpler case of a lattice scalar field with center. We have seen the
 Shannon term gives severe ambiguities to the entropy. Consider instead the mutual information between two regions $V$ and $W$, and let us call $A$ and $B$ to the set of points 
in the lattice on each of these regions, where we only take field and no momentum degree of freedom. The center is generated by the field in $A$ and $B$. The classical Shannon term 
in the entropy gets in the mutual information according to (\ref{mi}), (\ref{ffi}) and(\ref{ffi1})
\bea
&-&\int (\Pi_{a\in AB} d\phi_a) \, p_{AB}(\{\phi\}) \log(p_{AB}(\{\phi\})/(p_A(\{\phi\})p_B(\{\phi\})))\nonumber\\
&=& \frac{1}{2}\textrm{tr} \left[\log(X^A)+\log(X^B)-\log(X^{AB}))\right]\,,\label{got}
\eea
where $X^Z_{ab}=\langle \phi_a \phi_b\rangle$ is the field correlation matrix in a region $Z$. The combination (\ref{got}) is positive because the 
logarithm is an operator monotonic function \cite{petz}.

In contrast to the case of Shannon entropy, the classical mutual information and relative entropy have a well defined expression 
for continuum variables. This is because in discretizing a continuous variable $\phi$, two different probability distributions 
both decrease as $N^{-1}$  ($N$ the number of discrete points), and the $\log(N)$ terms cancel in the formula for the relative entropy. 
In other terms, while $\log(p(\phi))$ is a logarithm of a dimensionfull quantity, the probability density, the ratios appearing 
in the logarithm for relative entropy quantities, $\log(p_1(\phi)/p_2(\phi))$, do not have dimensions.  

Of course (\ref{got}) is only part of the mutual information between $V$ and $W$. The combination of this term with the quantum 
one in (\ref{mi}) must satisfy monotonicity. For example (\ref{got}) must be smaller than the full mutual information of 
regions $A$ and $B$ with all momentum and field variables in the algebras. Numerical evaluation in a two dimensional lattice 
shows the classical mutual information (\ref{got}) is in fact comparable to (and smaller) than the mutual information $I(A,B)$ 
of the full algebras on the same regions. 

However, if regions $A$ and $B$ are included in the thin boundary shells of $V$ and $W$, numerical value of this classical term will be small 
in the lattice. The same can be said for the classical terms of the mutual information or relative entropy for gauge theories. Because of monotonicity under inclusion of algebras, 
mutual information or relative entropy for these thin regions give an upper bound to the analogous classical quantities induced by any choice of  center. It is reasonable to expect that, for example, mutual information for thin objects at a fixed distance should vanish with width in the continuum limit.\footnote{We have checked this numerically for free fermion and scalar fields in the lattice in $d=2$.} Hence, the Shannon term should have no physical 
significance in the continuum limit since it does not contribute to relative entropy quantities. Interestingly, on the other hand, the averaged quantum relative entropies of (\ref{rerelala}) must give 
the same result independently of the choice of the center.


\section{Final comments}
Summarizing the results, the local entropy for gauge theories in the lattice is well defined as the entropy of the global state in a 
local algebra of gauge invariant operators. However, the connection of algebras to regions is subject to ambiguities which, for 
the special case of trivial center and entanglement entropy interpretation of the local entropy, are in correspondence to gauge fixings. These 
ambiguities cannot be avoided in the lattice but are not relevant to universal quantities in the continuum limit. 

We have focused on Abelian gauge fields. We do not expect changes to our main conclusions for the case of non Abelian gauge fields. In particular, algebras with trivial center can be formed by gauge fixing with maximal trees of links on the boundary. The magnetic center choice can also be formed in analogous manner taking only (mutually commuting) Wilson loops on the boundary. The gauge invariant electric generators require some modifications (see for example \cite{yaffe}), leading to a non commutative subalgebra. The expression of the integrated Gauss law is less transparent in this case. Hence, we do not know what is the natural generalization of the electric center choice in the non abelian case. It would also be of interest to see if the extended lattice construction is equivalent to the entropy of some particular choice of algebra for the model in the non abelian case. This is important to the physical relevance of this construction for the continuous limit.   

This work suggests the evaluation of local entropy in the continuum for gauge fields would follow the 
usual route of the replica trick to produce the traces of powers of the local density matrices. In order to have this local
 density matrices we should either choose a gauge fixing along the boundary of the region, or, in the case of non trivial center, 
 we should restrict the integration of the fields on the path integral to specific boundary conditions with some fields fixed on the boundary, and then average the entropy over the different values of these boundary fields. For example, we have a fixed normal
  electric field for the electric center choice. Barring boundary contributions, all these different prescriptions must give the 
  same results for finite universal terms. We left a more detailed derivation of the replica method in gauge theories for future work.    

We have found conflicting results for the topological entanglement entropy for a $Z_2$ model on the lattice based on the usual construction. Since topological theories do not have local degrees of freedom the entropy is specially 
 dependent on the choice of boundary details.  
Probably, the right way to define it requires the theory is not fully gapped, and there are some high energy local degree 
of freedom giving place to the topological model in the infrared. This allows us to take the continuum limit, and to compute 
the topological entropy as a term of the mutual information of a smooth region. The topological entropy contributes to the entropic 
c-function of the 3-dimensional c-theorem, which can be defined in this way, as the constant term ($\epsilon$ independent term) of the mutual 
information $I(R-\epsilon/2,R+\epsilon/2)$ for two concentric circles of radius $R+\epsilon/2$ and $R-\epsilon/2$ \cite{mutualc}, in the limit of small $\epsilon$.  
The necessity of UV degree of freedom in support of the topological model seems also natural from the fact that topological 
entropy is negative and then it needs and area term to compensate for the sign. 

\section*{Acknowledgements}

This work was supported by CONICET, CNEA
and Universidad Nacional de Cuyo, Argentina.

\end{document}